\documentclass[10pt,a4paper,twoside]{article}
\usepackage{epsfig}
\usepackage{baltlat6}
\usepackage{wrapfig}
\begin{document}
\ \ \vspace{-0.5mm}

\setcounter{page}{409}
\vspace{-2mm}

\titlehead{Baltic Astronomy, vol.\,16, 409--420, 2007}

\titleb{PHOTOMETRY OF STAR CLUSTERS IN THE M\,31 GALAXY.\\  APERTURE
SIZE EFFECTS}

\begin{authorl}
\authorb{D.~Narbutis}{1},
\authorb{V.~Vansevi\v{c}ius}{1},
\authorb{K.~Kodaira}{2},
\authorb{A.~Brid\v{z}ius}{1} and
\authorb{R.~Stonkut\.{e}}{1}
\end{authorl}

\begin{addressl}
\addressb{1}{Institute of Physics, Savanori\c{u} 231, Vilnius LT-02300,
Lithuania,\\ wladas@astro.lt}
\addressb{2}{The Graduate University for Advanced Studies (SOKENDAI),
Shonan Village, Hayama, Kanagawa 240-0193, Japan}
\end{addressl}

\submitb{Received 2007 November 30; accepted 2007 December 14}

\begin{summary} A study of aperture size effects on star cluster
photometry in crowded fields is presented.  Tests were performed on a
sample of 285 star cluster candidates in the South-West field of the
M\,31 galaxy disk, measured in the Local Group Galaxy Survey mosaic
images (Massey et al. 2006).  In the majority of cases the derived {\it
UBVRI} photometry errors represent the accuracy of cluster colors well,
however, for faint objects, residing in crowded environments,
uncertainties of colors could be underestimated.  Therefore, prior to
deriving cluster parameters via a comparison of measured colors with SSP
models, biases of colors, arising due to background crowding, must be
taken into account.  A comparison of our photometry data with {\it
Hubble Space Telescope} observations of the clusters by Krienke and
Hodge (2007) is provided.  \end{summary}


\resthead{Aperture size effects in star cluster photometry}
{D.~Narbutis, V.~Vansevi\v{c}ius, K.~Kodaira et al.}

\sectionb{1}{INTRODUCTION}

The accuracy of star cluster parameter quantification, based on a
comparison of colors, derived from {\it UBVRI} photometry with single
stellar population (SSP) models, has been tested by Narbutis et al.
(2007).  It was demonstrated, that an uncertainty of cluster colors less
than $\sim$\,$0.03$ mag is required  to achieve a reasonable
accuracy of age and extinction.  Therefore, to estimate the influence of
background contaminants on the accuracy of object colors, we studied
aperture size effects, based on a sample of 285 star cluster candidates
in the South-West field of the M\,31 galaxy published by Narbutis et al.
(2008).

The aperture photometry errors of extended objects are determined by
photon noise, photometric background uncertainty and calibration
procedure errors.  However, in the case of extragalactic star clusters
projected on a crowded galaxy disk (e.g., M\,31), the flux from
contaminating stars biases true colors of the objects, and in some cases
the contamination effect exceeds photometric errors.  The colors and
density of dominating contaminants depend mainly on the age of
background stellar populations and on the object location in the galaxy.
In general, background stars are distributed randomly, therefore, a
probability of their contaminating effect increases with the aperture
size.

The resolution of ground-based images is limited by the atmospheric
seeing.  Therefore, colors of star clusters are affected by
contaminating objects to a degree which depends on the quality of
images, characterized by the full width at half maximum (FWHM) of the
point spread function (PSF), and the location of the contaminant in
respect to the aperture.  An alteration of cluster's true colors also
depends on the contaminant and the object luminosity ratio -- faint
star clusters are more affected.

Contaminants, residing outside the aperture, can be partly avoided by
using high resolution imaging, e.g., with the {\it Hubble Space
Telescope} (HST), however, the sky background determination becomes
more complicated in this case (see discussion by Krienke \& Hodge 2007).
In order to estimate the resolution effect, we performed a comparison of
the cross-identified cluster photometry from our sample and from the
sample studied by Krienke \& Hodge (2007).

\sectionb{2}{THE DATA}

An initial survey of star clusters in the South-West field of the M\,31
disk was conducted by Kodaira et al.  (2004) in an area of about 500
square arcminutes, making use of the high-resolution imaging capability
of the Subaru Suprime-Cam (Miyazaki et al. 2002), and a catalog of star
cluster candidates up to $V\,\sim\,19.0$ mag was presented.  The study
of the structural parameters of this sample was based on the $V$-band
Suprime-Cam image analysis (\v{S}ablevi\v{c}i\={u}t\.{e} et al. 2006,
2007), showing a wide range of the intrinsic object sizes.  The {\it
UBVRI} ($R$ and $I$ passbands being in the Cousins system) photometry of
these clusters was performed on the Local Group Galaxy Survey (LGGS;
Massey et al. 2006) images by Narbutis et al. (2006).

A sample of objects has been extended up to $V\!\sim\!20.5$ mag, and the
catalog of 285 star cluster candidates, supplemented with multi-band
color maps combined from the ultra-violet, optical, infra-red and radio
images, was provided by Narbutis et al.  (2008).  The photometry
presented in this study was performed on the LGGS mosaic images through
individually selected circular apertures by employing the XGPHOT package
in the IRAF program system (Tody 1993).  The shape of mosaic image PSFs
(scale is 0.27\arcsec/pix) was homogenized to the resolution of
FWHM\,=\,1.5\arcsec, and was used for photometry.  This procedure
ensures a constant aperture correction for an object measured in all
passbands.  A typical aperture diameter was set to 3\arcsec\ for the
majority of clusters, and was extended up to 10\arcsec\ for the
brightest objects.  The photometric background was determined in
individually selected and object-centered circular annuli with typical
inner and outer diameters of 6\arcsec\ and 16\arcsec, respectively.  For
some objects, residing on the largely variable background, the circular
background determination zones were selected individually in
representative areas.

Each object was measured in two or three different LGGS fields.  The
$V$-band magnitude and colors were transformed to the standard system
using the prescription given by Massey et al.  (2006).  The reliability
of the calibrations was discussed by Narbutis, Stonkut\.{e} and
Vansevi\v{c}ius (2006).  $V$-band magnitudes and colors of each object,
measured in different fields, were examined interactively, and their
weighted averages were calculated.  The photometric errors provided in
the catalog account for the photon noise and the errors of calibration
procedure.  We note that $\sim$\,10\% of the objects from our sample
have suspected variable stars contaminating the aperture photometry
results.  They were revealed by comparing overlapping LGGS fields
observed on different nights.

\vbox{\centerline{\psfig{figure=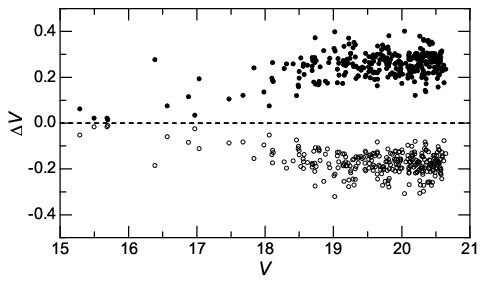,width=120truemm,angle=0,clip=}}
\vspace{0.5mm}
\captionb{1}{Aperture size test results of the 285 star cluster
candidates.  $\Delta V$ is the difference of $V$ magnitudes measured
through the apertures increased or decreased by 0.6\arcsec\ in respect
to the the main aperture (open/filled circles, respectively) and $V$
magnitudes, measured trough the main aperture.  The aperture diameter of
3\arcsec\ was adopted for the majority of clusters and up to 10\arcsec\
for the brightest objects.}}

\subsectionb{2.1}{Aperture size test}

In order to study  aperture size effects, arising due to
contamination by background stars, on the accuracy of colors, all star
cluster candidates were additionally measured trough the circular
apertures increased/decreased by 0.6\arcsec\ in respect to the main
aperture diameter.  The same photometry reduction procedure, as that
applied for the main aperture data, was used, and the differences of the
$V$ magnitudes and colors, measured through the increased or decreased
and the main apertures, were calculated.  These results also provide an
estimate of possible star cluster color bias, arising due to aperture
centering and size mismatch when measured in images from different
surveys.

In Figure 1 we plot $\Delta V$ versus $V$. Here $\Delta V$ are
differences of $V$ magnitudes measured through the apertures increased
or decreased by 0.6\arcsec\ and $V$ magnitudes obtained trough the main
aperture.  Since bright objects were measured through large main
apertures, that resulted in small magnitude differences.  However, for
faint objects ($V\!>\!18.5$ mag) $\Delta V$ is significantly larger and
has to be taken into account for the star cluster mass estimate.

In Figure 2 we plot $\Delta(B\!-\!V)$, $\Delta(U\!-\!B)$ and
$\Delta(V\!-\!I)$ versus $V$ magnitudes.  The distribution of
$\Delta(B\!-\!V)$ is symmetric.  However, for faint objects
($V\!>\!18.0$ mag) the distributions of $\Delta(U\!-\!B)$ and
$\Delta(V\!-\!I)$ are asymmetric, suggesting that some objects are
affected by blue and red contaminants, respectively.  The same color
differences are plotted versus $B\!-\!V$ in Figure 3. $\Delta(U\!-\!B)$
shows an asymmetric distribution due to the influence of blue
contaminants, which is stronger for star cluster candidates with
$B\!-\!V\!>\!0.5$.  $\Delta(V\!-\!I)$ shows an asymmetric distribution
for the objects with $B\!-\!V\!<\!0.75$ due to the influence of red
contaminants.

Uncertainties of cluster colors, arising due to the influence of
contaminants,

\vbox{\centerline{\psfig{figure=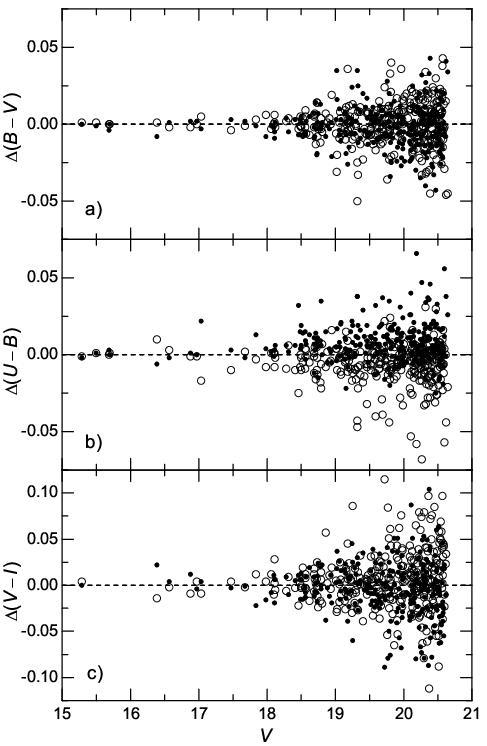,width=120truemm,angle=0,clip=}}
\vspace{0.5mm}
\captionb{2}{The same as in Figure 1, but for the $\Delta(B\!-\!V)$,
$\Delta(U\!-\!B)$ and $\Delta(V\!-\!I)$ color differences.}}

\newpage
\vbox{\centerline{\psfig{figure=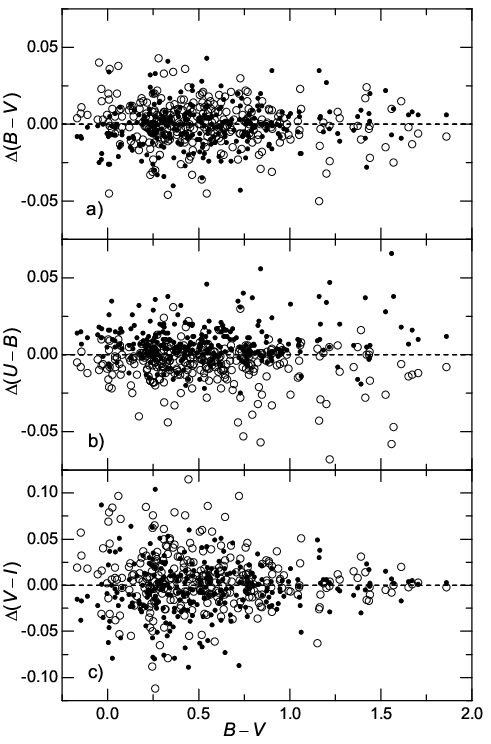,width=120truemm,angle=0,clip=}}
\vspace{0.5mm}
\captionb{3}{The same as in Figure 2, but the $\Delta(B\!-\!V)$,
$\Delta(U\!-\!B)$ and $\Delta(V\!-\!I)$ color differences are plotted
versus $B\!-\!V$.}}

\noindent
were estimated for each object based on $\Delta(U\!-\!B)$,
$\Delta(B\!-\!V)$ and $\Delta(V\!-\!I)$.  $U\!-\!B$ versus $B\!-\!V$
diagrams of 285 star cluster candidates are plotted in Figure 4. The
error bars indicate photometric errors, $\sigma(U\!-\!B)$,
$\sigma(B\!-\!V)$ (panel a), and $\Delta(U\!-\!B)$, $\Delta(B\!-\!V)$
(panel b).  Due to the influence of contaminants for a fraction of
objects, 4\% for $U\!-\!B$ and 7\% for $B\!-\!V$, the $\Delta$ values
exceed the $\sigma$ values.  In Figure 5 the corresponding $V\!-\!I$
versus $B\!-\!V$ diagrams are plotted.  The $\Delta(V\!-\!I)$ values
exceed the $\sigma(V\!-\!I)$ values for 21\% of star clusters.

In the majority of cases photometric errors represent a true uncertainty
of cluster colors well, however, for some faint objects, residing in
crowded environments, these errors can be largely underestimated.  It is
noteworthy to stress that possible biases of colors have to be taken
into consideration prior to the cluster parameter quantification
procedure, based on the SSP models.

\subsectionb{2.2}{Comparison with $HST$ data}

A large sample of star clusters in M\,31 has been studied recently with
HST by Krienke \& Hodge (2007).  They provide a catalog of {\it BVI}
photometry (transformed from the HST to the Johnson-Cousins system) for
343 star clusters detected in the Wide Field and Planetary Camera 2
(WFPC2) images.  Krienke \& Hodge (2007) measured integrated magnitudes
of star clusters through individual isophotal apertures.  Note, that the
aperture size distribution, provided in their Fig.~7, peaks at
$\sim$\,3\arcsec\ and is equal to a typical aperture diameter used for
photometry of our cluster sample.

Eight HST data sets, used by Krienke \& Hodge (2007) and overlapping
with our field, were obtained from the Multimission Archive at the Space
Telescope Science Institute (MAST)\footnote{~See
http://archive.stsci.edu/hst/search.php}, and used for object
identification and analysis of their environment.

In total, 36 common objects were cross-identified in the studies by
Krienke \& Hodge (2007) and Narbutis et al.  (2008).  Differences of $V$
(34 objects), $B\!-\!V$ (14) and $V\!-\!I$ (25) are shown in Figure 6
versus $V$ magnitude.  The error bars indicate photometric errors from
Krienke \& Hodge (2007); for objects without provided errors we assumed
representative errors of $V$ magnitude.  Faint objects show a larger
magnitude and color difference scatter.  Two bright objects with
$V\!\sim\!14.3$ mag and $V\!\sim\!15.7$ mag, not shown in Figure 6, have
$V$ magnitude differences of 0.08 and 0.03 mag, respectively.  Four
objects, all residing in one HST field, have a significant $V$ magnitude
difference, presumably due to a zero-point inaccuracy.  However, their
colors do not show significant discrepancies.  A much smaller aperture
size was used in our study for the semi-resolved star cluster indicated
by a filled triangle, however, the measured $B\!-\!V$ color matches well
between both studies.

Color differences are shown as a function of colors in Figure 7.
Vertical and horizontal error bars show photometric errors from Krienke
\& Hodge (2007) and Narbutis et al.  (2008), respectively.  The
$B\!-\!V$ colors are mostly affected by blue contaminants and the
$V\!-\!I$ colors -- by red contaminants due to differing aperture
centers, shapes and sizes.  Different photometric background
determination strategies used in both studies might also be responsible
for the observed scatter.  Note, that four objects, shown in panel (a)
with $B\!-\!V\!\sim\!0.3$, all reside in one HST field.  Therefore, the
difference of colors could be of systematic origin.  Variable stars

\vbox{\centerline{\psfig{figure=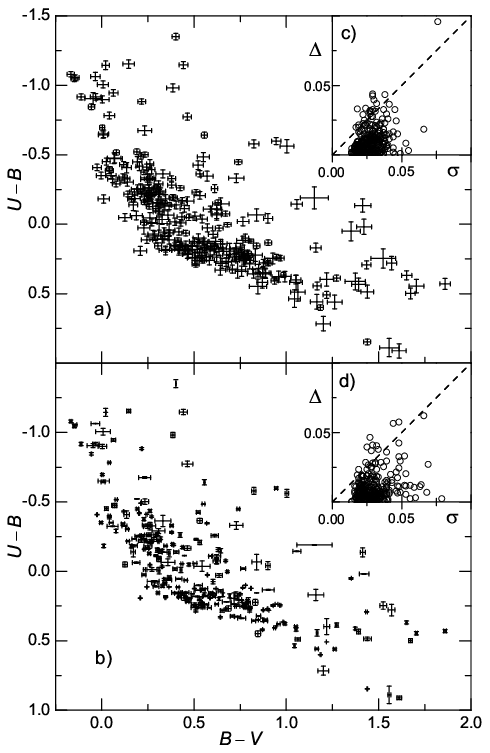,width=110truemm,angle=0,clip=}}
\vspace{0.5mm}
\captionb{4}{The $U\!-\!B$ vs. $B\!-\!V$ diagrams for the 285 star
cluster candidates. Error bars indicate photometric errors, $\sigma$,
and the aperture size effects, $\Delta$, in panels (a) and (b),
respectively. In the insets $\Delta(B\!-\!V)$ of the aperture size
test vs. photometric error, $\sigma(B\!-\!V)$, and $\Delta(U\!-\!B)$
vs. $\sigma(U\!-\!B)$ are shown in panels (c) and (d), respectively.}}

\newpage
\vbox{\centerline{\psfig{figure=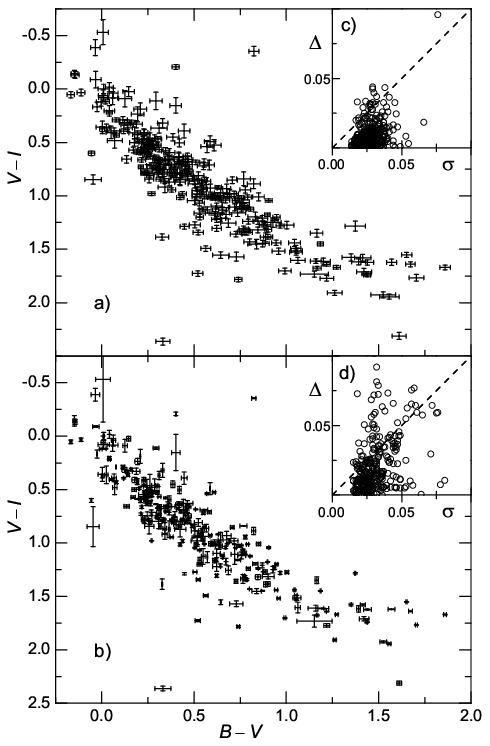,width=110truemm,angle=0,clip=}}
\vspace{0.5mm}
\captionb{5}{The same as in Figure 4, but $V\!-\!I$ vs. $B\!-\!V$
diagrams are plotted. Panel (c) is the same as in Figure 4; in panel (d)
$\Delta(V\!-\!I)$ vs. $\sigma(V\!-\!I)$ is shown.}}

\newpage
\vbox{\centerline{\psfig{figure=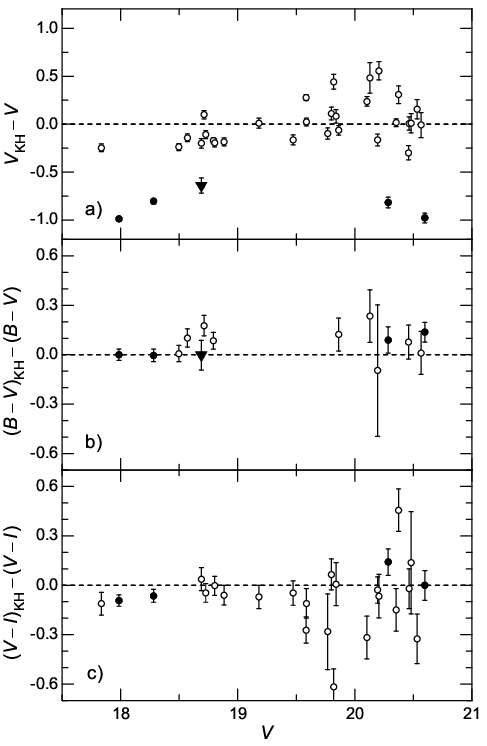,width=110truemm,angle=0,clip=}}
\vspace{0.5mm}
\captionb{6}{Differences of $V$ magnitudes and colors from Krienke \&
Hodge (2007) (marked with KH) and Narbutis et al.  (2008) are shown for
34, 14 and 25 clusters in panels (a), (b) and (c), respectively.  Error
bars indicate photometric errors from Krienke \& Hodge (2007).  The
deviations of $V$ magnitudes of four objects (filled circles), residing
in the same HST field, are presumably subject to zero-point inaccuracy.
However, this does not affect their colors.  A significantly smaller
aperture size was applied in our study for a cluster marked by a filled
triangle.}}

\newpage
\vbox{\centerline{\psfig{figure=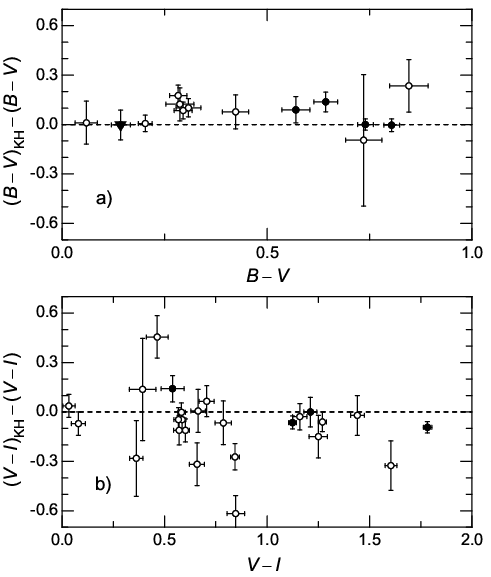,width=110truemm,angle=0,clip=}}
\vspace{0.5mm}
\captionb{7}{The same as in Figure 6, but color differences are plotted
vs. corresponding colors.  Vertical and horizontal error bars show
photometric errors from Krienke \& Hodge (2007) and Narbutis et al.
(2008), respectively.  Due to differing aperture centering, shapes and
sizes used in both studies, the $B\!-\!V$ colors are predominantly
affected by blue contaminants, and $V\!-\!I$ colors by red contaminants.
Note, that four objects making a compact group in panel (a),
$B\!-\!V\!\sim\!0.3$, reside in one HST field.  For description of the
objects, shown by filled circles and a filled triangle, see Figure 6.}}

\vskip5mm

\noindent
residing close to the studied objects might contaminate aperture
magnitudes and colors, when observations from multiple epochs are
compared.  However, objects with the smallest error bars show a
reasonably good match between LGGS and HST photometry data over a wide
range of star cluster colors.

\newpage

\sectionb{3}{SUMMARY}

In order to study crowding effects on star cluster aperture photometry,
we performed an aperture size test for a sample of 285 star cluster
candidates in the South-West field of the M\,31 galaxy (Narbutis et al.
2008).  The influence of background contaminants was estimated by
measuring cluster colors through additional apertures increased or
decreased by 0.6\arcsec\ in respect to the main aperture diameter
(typically 3\arcsec).  Photometric errors for 4\% (for $U\!-\!B$), 7\%
(for $B\!-\!V$) and 21\% (for $V\!-\!I$) of clusters are smaller than
the errors arising due to the influence of contaminating background
stars.  The $U\!-\!B$ color of red objects tends to be systematically
bluer (contaminated by blue stars), and an opposite effect is observed
for $V\!-\!I$ of blue objects (contaminated by red stars).

We cross-identified 36 objects from our sample (LGGS) with the star
clusters from Krienke \& Hodge (2007) (HST).  Despite different
apertures (sizes, shapes and centering), photometric background
determination and photometric calibration strategies used in both
studies, $V$ magnitudes and $B\!-\!V$, $V\!-\!I$ colors show a
reasonably good agreement.

The aperture size test allows us to identify star clusters which have
integrated colors affected by background contaminants.  The influence of
background objects can be eliminated (to some degree) by measuring
clusters trough individually selected isophotal apertures on high
resolution images (e.g.,  HST; Krienke \& Hodge 2007).  However, the
influence of contaminants, projected on star clusters within the
applied aperture, can be estimated only statistically.  Using
ground-based (lower resolution) images for aperture photometry of faint
star clusters it is difficult to avoid a strong contamination of colors
by neighboring stars.  Therefore, biases of the object colors have to be
accounted for prior to deriving cluster parameters via a comparison of
the observed colors with the SSP model data.

\vskip5mm

ACKNOWLEDGMENTS. This work was financially supported in part by a Grant
of the Lithuanian State Science and Studies Foundation. The star cluster
survey is based on the Suprime-Cam images, collected at the Subaru
Telescope, which is operated by the National Astronomical Observatory
of Japan. The data presented in this paper were partly obtained from
the Multimission Archive at the Space Telescope Science Institute.
This research has made use of SAOImage DS9, developed by Smithsonian
Astrophysical Observatory.

\vskip5mm

\References

\refb Kodaira~K., Vansevi\v{c}ius~V., Brid\v{z}ius~A., Komiyama~Y.,
Miyazaki~S., Stonkut\.{e}~R., \v{S}ablevi\v{c}i\={u}t\.{e}~I., Narbutis~D. 2004, PASJ, 56, 1025

\refb Krienke~O. K., Hodge~P. W. 2007, PASP, 119, 7

\refb Massey~P., Olsen~K.\,A.\,G., Hodge~P.\,W., Strong~S.\,B., Jacoby~G.\,H., Schlingman~W., Smith~R.\,C. 2006, AJ, 131, 2478

\refb Miyazaki~S. et al. 2002, PASJ, 54, 833

\refb Narbutis D., Brid\v zius A., Stonkut\.e V., Vansevi\v cius V.
2007, Baltic Astronomy, 16, 421 (this issue)

\refb Narbutis~D., Vansevi\v{c}ius~V., Kodaira~K.,
\v{S}ablevi\v{c}i\={u}t\.{e}~I., Stonkut\.{e}~R., Brid\v{z}ius~A. 2006,
Baltic Astronomy, 15, 461

\refb Narbutis~D., Stonkut\.{e}~R., Vansevi\v{c}ius~V. 2006, Baltic
Astronomy, 15, 471

\refb Narbutis~D., Vansevi\v{c}ius~V., Kodaira~K., Brid\v{z}ius~A.,
Stonkut\.{e}~R. 2008, ApJS, submitted

\refb \v{S}ablevi\v{c}i\={u}t\.{e}~I., Vansevi\v{c}ius~V., Kodaira~K.,
Narbutis~D., Stonkut\.{e}~R., Brid\v{z}ius~A. 2006, Baltic Astronomy,
15, 547

\refb \v{S}ablevi\v{c}i\={u}t\.{e}~I., Vansevi\v{c}ius~V., Kodaira~K.,
Narbutis~D., Stonkut\.{e}~R., Brid\v{z}ius~A. 2007, Baltic Astronomy,
16, 397 (this issue)

\refb Tody~D. 1993, {\it Astronomical Data Analysis Software and Systems II}, 52, 173

\end{document}